# Light-induced magnetization driven by interorbital charge motion in a spin-orbit assisted Mott insulator α-RuCl$_3$

T. Amano[1], Y. Kawakami[1], H. Itoh[1], K. Konno[1], Y. Hasegawa[1], T. Aoyama[1], Y. Imai[1], K. Ohgushi[1], Y. Takeuchi[1], Y. Wakabayashi[1], K. Goto[2], Y. Nakamura[2], H. Kishida[2], K. Yonemitsu[3], and S. Iwai[1*]

[1]Department of Physics, Tohoku University, Sendai 980-8578, Japan

[2] Department of Applied Physics, Nagoya University, Nagoya 464-8603 Japan

[3]Department of Physics, Chuo University, Tokyo 112-8551, Japan

 71.27.+a, 74.25.Gz, 78.47.J-

*s-iwai@tohoku.ac.jp

Abstract

In a honeycomb-lattice spin-orbit assisted Mott insulator $\alpha$-RuCl$_3$, an ultrafast magnetization is induced by circularly polarized excitation below the Mott gap. Photo-carriers play an important role, which are generated by turning down the synergy of the on-site Coulomb interaction and the spin-orbit interaction realizing the insulator state. An ultrafast 6-fs measurement of photo-carrier dynamics and a quantum mechanical analysis clarify the mechanism, according to which the magnetization emerges from a coherent charge motion between different $t_{2g}$ orbitals ($d_{yz}$-$d_{xz}$-$d_{xy}$) of Ru$^{3+}$ ions. This ultrafast magnetization is weakened in the antiferromagnetic (AF) phase, which is opposite to the general tendency that the inverse Faraday effect is larger in AF compounds than in paramagnetic ones. This temperature dependence indicates that the interorbital charge motion is affected by pseudo-spin rotational symmetry breaking in the AF phase.

Emergent properties of strongly correlated charges and spins enable us to expect a new pathway for light-induced spatial/time reversal symmetry breaking [1-6]. In particular, spin-orbit assisted Mott insulators exhibiting quantum frustrations [7-10] are the promising candidate for realizing the light-induced magnetization (or equivalently the time reversal symmetry breaking) because of their charge-orbital-spin entanglement. Note that the materials for light-induced magnetization have been discussed mainly in anti-ferromagnets or weak ferromagnets [11-14].

A layered insulator α-RuCl$_3$ with honeycomb lattice (left side of Fig. 1(a)) has attracted much attention as a spin-orbit assisted Mott insulator [8-10, 15-19]. The strong spin-orbit interaction (SOI) splits the $t_{2g}$ states of a Ru$^{3+}$ ion into pseudospins $J_{\text{eff}}$=1/2 and $J_{\text{eff}}$=3/2. Thus, the Mott insulating state of this compound is realized by the synergy of the on-site Coulomb interaction $U$ and the SOI [20]. In the edge-sharing octahedron (right side of Fig. 1(a)), no long-range order is allowed if only the bond-dependent anisotropic exchange interaction (Kitaev interaction) exists between the $J_{\text{eff}}$=1/2 pseudospins on the honeycomb lattice. In fact, magnetic orders are not detected down to $T_N$, although a zigzag antiferromagnetic (AF) order in the honeycomb plane (ab-plane) is reported below $T_N$=7 K-14 K which depends on the volume fractions of $R\bar{3}$ and $C2/m$ (Supplementary 1 [21])[22-30]. Here, intersite hopping $t$ between different $t_{2g}$ orbitals (such as $d_{xz} - d_{yz}$) of Ru ions is essential. Because of this characteristic intersite hopping, peculiar charge and spin dynamics on the ultrafast time scale are expected. Another important characteristic is in-gap states below the Mott-Hubbard transition ($E_{\text{gap}}$=1.2 eV below 50 K [31-35]) (Fig. 1(b)) that were discussed in terms of orbital excitations influenced by the SOI. They are recently assigned to excitations from $J_{\text{eff}}$=3/2 to $J_{\text{eff}}$=1/2 states (lower inset of Fig. 1(b)) [32-34]. The in-gap states have a broad and overlapped $\varepsilon_2$ spectrum in the range 0.2-0.9 eV (Fig. 1(b). The three peaks (0.31, 0.53, and 0.72 eV) roughly reflect different numbers of $J_{\text{eff}}$=3/2 holes substituting for $J_{\text{eff}}$=1/2 holes in the Mott insulating background with one $J_{\text{eff}}$=1/2 hole at each site [32]. Because of the spin-orbit assisted nature of this Mott insulator [20], an excitation of these spin-orbit coupled (SOC) in-gap states in α-RuCl$_3$ may take advantage of charge degrees of freedom in light-induced phenomena.

In this letter, we report that a circularly polarized excitation below the Mott gap produces magnetization perpendicular to the honeycomb plane of α-RuCl$_3$. Ultrafast dynamics of photo-carriers and a quantum many-body analysis show that the ultrafast magnetization is driven by the hopping $t$ between different $t_{2g}$ orbitals ($d_{yz}$-$d_{xz}$, $d_{xz}$-$d_{xy}$, $d_{xy}$-$d_{yz}$) as shown in Fig. 4(a).

The samples used in this study were prepared by the chemical vapor transport method [35]. We performed transient polarization rotation ($\Delta\theta$) and transmission ($\Delta T/T$) measurements for circularly polarized excitation using the pump-probe method with the time resolution of 100 fs. The probing spectral range covers 0.5-1.1 eV for the pump light with 0.30, 0.62, 0.89 and 1.55 eV. $\Delta R/R$ spectra are measured for excitations of 0.62 and 0.89 eV. In addition, we measure $\Delta R/R$ with higher time-resolution by utilizing a carrier-envelope phase (CEP) stabilized 6-fs near infrared pulse for investigating ultrafast charge coherence [36, 37].

The upper panel of Fig. 2(a) shows time evolutions of $\Delta\theta$ (probe energy $E_{pr}$=0.62 eV, ||ab plane) induced by the circularly polarized pump pulse (pump energy $E_{pu}$=0.89 eV, intensity $I_{ex}$=4 mJ/cm$^2$) at 4 K (dashed curves, $T < T_N$) and 17 K (solid curves, $T > T_N$). The helicity ($\sigma^+$, $\sigma^-$) of the pump pulse and the direction of the rotation angle are defined as those in the right-handed system (upper inset of Fig. 1(b)). The helicity-dependent responses as large as 4.5-5.5 degrees within the time resolution (ca. 100 fs) in Fig. 2(a) is attributed to the Faraday rotation reflecting an ultrafast magnetization perpendicular to the ab-plane ($\perp$ab) at both 4 K and 17 K (Supplementary 2 [21])[38]. The observation of such large $\Delta\theta$ without any long-range order at 17 K contrasts with the ultrafast inverse-Faraday effect in AF compounds [11-14]. The transient transmittance ($\Delta T/T$) in the lower panel of Fig. 2(a) shows concomitant charge dynamics with a finite decay time indicating real excitation of in-gap states. In fact, $\Delta\theta$ normalized by the thickness (ca. 50 μm) and $I_{ex}$ (4 mJ/cm$^2$) is 20 times larger than that of a prototypical AF magnet NiO (111) (1.14 deg. for $E_{pu}$=0.97 eV, $E_{pr}$=1.59 eV, $I_{ex}$= 10 mJ/cm$^2$, thickness 100 μm [14]). It is 400 times larger than that of a paramagnet Terbium Gallium Garnet (0.15 deg. for $E_{pu}$=$E_{pr}$=1.55 eV, $I_{ex}$=3.1 mJ/cm$^2$, thickness 1 mm [39]).

The temperature dependence of $\Delta\theta$ at time delay ($t_d$) of 0 ps in the upper panel of Fig. 2(b) shows an abrupt change near $T_N$=7 K. The reduction of $\Delta\theta$ (of ca. 15 %) below $T_N$ indicates that the ultrafast magnetization ($\perp$ab) is weakened in the AF phase, which is opposite to the general tendency that the inverse Faraday effect is larger in AF compounds than in paramagnetic ones.

Above 20 K, $\Delta\theta$ is insensitive to the temperature (Supplementary 3 [21]).

The helicity-independent slower component of $\Delta\theta$ (x10 for $t_d$>1 ps in Fig. 2(a)) is attributed to a melting of the AF order (Supplementary 4 [21])[40]. In fact, $\Delta\theta$ at 10 ps drops near $T_N$ (lower panel of Fig. 2(b)). In contrast to the temperature sensitive nature of $\Delta\theta$, $\Delta T/T$ do not indicate any temperature dependence near $T_N$ (as shown in the lower panel of Fig. 2(a)), showing that the temperature dependence of $\Delta\theta$ is not attributed to a change in the steady-state absorption coefficient ($\alpha$) at $E_{pu}$. A large $\Delta T/T$ ~0.6 ($\Delta\alpha/\alpha$ of 0.3) reflects excited carrier dynamics which will be discussed later along with $\Delta R/R$ around the Mott Hubbard band.

To investigate a mechanism of light induced magnetization under the excitation below the Mott gap, we measure opto-magneto spectra ($\Delta\theta(E_{pr})$: rotation angle and $\Delta\eta(E_{pr})$: ellipticity) by changing $E_{pu}$. Figures 2(c) and 2(d) show the $\Delta\theta(E_{pr})$ and $\Delta\eta(E_{pr})$ spectra, respectively, for (i) $E_{pu}$ =0.30 eV, (ii) $E_{pu}$ =0.62 eV, and (iii) $E_{pu}$ =0.89 eV ($I_{ex}$=1.0 mJ/cm$^2$). As shown in (i) and (ii) of Figs 2(c) and 2(d), the energy where $\Delta\theta = 0$ ($\Delta\theta = 0$ intersection: green arrow) is equal to the energy where $\Delta\eta$ has a peak ($\Delta\eta$ peak: orange arrow), and it is ca. 0.9 eV for $E_{pu}$=0.30 eV, and ca. 0.6 eV for $E_{pu}$=0.62 eV. The $\Delta\theta = 0$ intersection and the $\Delta\eta$ peak correspond to the resonance energy $E_{res}$ in the magneto-optical spectrum $\theta = \frac{\omega l}{2cn} Im(\varepsilon_{xy})$ and $\eta = -\frac{\omega l}{2cn} Re(\varepsilon_{xy})$ ($\varepsilon_{xy}$: off-diagonal component of the permittivity, $n$: refractive index, and $l$: sample thickness) for the single resonance oscillator [41]. The red dots in the upper panel of Fig. 2(c) show the maximum values in the respective $\Delta\theta$ spectra (i.e., the excitation spectrum of $\Delta\theta$) for $E_{pu}$=0.30-0.89 eV. Since a probe light with energy higher than 0.9 eV is not transmitted, a true maximum cannot be detected for a pump light with or lower than 0.3 eV. Thus, the excitation spectrum of $\Delta\theta$ is underestimated at $E_{pu}$= 0.3 eV. Even in this limited situation, the tendency of the excitation spectrum in the upper panel of Fig. 2(c) is similar to that of $\varepsilon_2$ in the sense that $\Delta\theta$ becomes large for lower excitation energies. The ultrafast magnetization is absent for the excitation above the Mott gap (Supplementary 5 [21]). It is noteworthy that $E_{pu}+E_{res} \cong E_{gap}$=1.2 eV is satisfied for $E_{pu}$=0.30 and 0.62 eV. This relation suggests that a third-order nonlinear ($\chi^{(3)}$) process ($E_{pu}+E_{pr} = E_{gap}$) through a virtual excitation by pump light may partially contribute, although photocarriers are generated.

These results show that the probe pulse further excites the in-gap states during their lifetime (< 100 fs) to a state just above the Mott gap. This lifetime is comparable with the inverse of the bandwidth (~100 meV) of each peak in the $\varepsilon_2$ spectrum (Fig. 1(b)) and is supported by the results of the ultrafast measurement using a 6 fs pulse (Fig. 3(e)). The large $\Delta\theta$ in α-RuCl$_3$ is characteristic of the real excitation below the Mott gap and the 2-step excitation at $E_{pu}+E_{pr}$ which is resonant to the Mott gap.

The insulating state is affected by the excitation below the Mott gap, as shown in the transient reflectivity ($\Delta R/R$) spectra for $E_{pu}$=0.6 eV (Fig. 3(b)) and 0.89 eV (Fig. 3 (c)). The steady-state reflectivity ($R$) is shown in Fig. 3(a) for reference. The bleaching near the Mott-Hubbard transition at 1.2 eV is clearly detected, where the two-photon (2 x $E_{pu}$) absorption of the pump pulse is ruled out as the origin of the bleaching because of the linear dependence of $\Delta R/R$ on $I_{ex}$ (insets of Figs. 3(b) and 3(c)). This fact shows that photo-carriers are generated by the excitation below the Mott gap as well as the excitation above the Mott gap (black circles with dashed curve in Fig. 3(c)) [40, 42] (Supplementary 6 [21]). This is consistent with the fact that the Mott insulating state of this compound is realized by the synergy of the on-site Coulomb interaction $U$ and the SOI [20] (Supplementary 7[21]). Analyses of the time profiles of $\Delta R/R$ (Fig. 3(d)) by using a multi-exponential function (Supplementary 8[21]) show that the components with the decay time constants of < 0.1ps (50-60 %) and 0.25 ps (~35%) are dominant. They are consistent with the time constants of $\Delta T/T$ (Fig. 2(a)) confirming that the time profiles of $\Delta T/T$ reflect the photo-carrier dynamics (Supplementary 8[21]).

The fast (<0.1 ps) component of the recovery of $\Delta R/R$ is comparable with that of $\Delta \theta$ in Figs. 2(a) and 2(b). This fact indicates that the ultrafast magnetization is induced by the photo-carriers. The early-stage dynamics of the photo-carriers are captured by the $\Delta R/R$ measurement utilizing a 6-fs pulse ($E_{pu}$=0.55-0.95 eV, $E_{pr}$= 0.60 eV)[36], although a coherent polarization effect observed at $t_d$=-0.05-0.05 ps is non-essential to the photocarrier dynamics. In addition to the exponential function with a time constant of 0.17 ps (red line) which is roughly equal to those in Fig. 3(d)(0.25 ps), the time profile of $\Delta R/R$ in Fig. 3(e) has an oscillating component with a period of 36 fs as shown by the green circles in the inset (Supplementary 9[21]). Since the oscillating period (36 fs=h/(0.11 eV)) corresponds to the energies of the intersite hopping (0.1-0.2eV)) [43-45], the oscillation would be related to the intersite hopping processes including the interorbital one. Note that phonon energies are restricted to a range below 40 meV [15-17, 34]. Such coherent charge motions are often realized by the simultaneous application of a light-field force to many electrons that are correlated by the Coulomb interaction [46, 47] (Supplementary 9[21]). The dephasing time of the oscillation (60 fs, (Supplementary 9[21]) reflecting the coherence time of the intersite charge motion is approximately equal to the decay time of $\Delta\theta$. Therefore, the ultrafast magnetization is related to the coherent charge motion and its coherence time is consistent with the spectral bandwidths of SOC in-gap states (~100 meV). This oscillation survives even at 300 K as shown by the blue circles in the inset of Fig. 3(e). This is consistent with the temperature-insensitive $\Delta\theta$ up to 300 K (Supplementary 3[21], Fig. S3)

Here, we theoretically study photo-induced dynamics of the spin and orbital degrees of freedom to analyse the mechanism for the photo-induced magnetization in α-RuCl$_3$. We consider a Hubbard model consisting of $d_{yz}$, $d_{xz}$, and $d_{xy}$ orbitals on each site of the honeycomb lattice [43-45] shown in Fig. 4(a) (Supplementary 10[21]) [48, 49]. Figure 4(b) shows calculated linear optical conductivity (σ) spectra for ||a and ||b. Below the Mott gap of ca. 1 eV, some infrared peaks are found in the spectra. They are attributed to excitations from $J_{eff}$=3/2 to $J_{eff}$=1/2 states [32-34] (Supplementary 10[21]). Because of the finite-size effect, the excitation spectra are quite discrete even above the Mott gap. The pseudospin dynamics are calculated on the basis of numerical solutions to the time-dependent Schrödinger equation. Photo-excitation by a circularly polarized pulse is introduced through the Peierls phase (Supplementary 10 [21]) [48, 49]. The magnetization $M_{Ph}$ (Fig. 4(c)) in the direction of ⊥ab is given by $M_{\text{Ph}}= -2j^{(1/2)}_\perp$ where the pseudospin density $j^{(1/2)}$ is defined in Supplementary 10. The helicity-dependent $M_{\text{ph}}$ grows during the excitation of 96 fs (= pulse width) with ℏω=0.3 eV or 0.6 eV showing that the magnetization arises within the pulse width (where the direction of $M_{\text{ph}}$ is from the back to the front of Fig. 4(a) for σ⁺).

To interpret the photo-induced dynamics, we use a high-frequency expansion in the framework of quantum Floquet theory to evaluate an effective magnetic field. This effective field emerges from the commutators among the kinetic operators on the three bonds (Supplementary 11[21]) [50]. Thus, the charge hopping between different $t_{2g}$ orbitals ($d_{yz}$-$d_{xz}$-$d_{xy}$) (Fig. 4(a)) is essential to the photo-induced magnetization. The reduction of the ultrafast magnetization below $T_N$ indicates that the interorbital charge motion is affected by pseudo-spin rotational symmetry breaking in the AF phase. The coherent charge response in Fig. 3(e) shows a very important perspective for coherent control of the light-induced magnetization.

In summary, this article demonstrates the ultrafast magnetization in the multiorbital electron system α-RuCl$_3$ by measuring the helicity-dependent polarization rotation. The light-induced magnetization is induced by the charge motion between different $t_{2g}$ orbitals.

### Acknowledgements

This work was supported by JST CREST (JPMJCR1901), MEXT Q-LEAP (JPMXS0118067426).

References

[1] H. Aoki, N. Tsuji, M. Eckstein, M. Kollar, T. Oka, & P. Werner, Nonequilibrium dynamical mean-field theory and its applications. *Rev. Mod. Phys.* **86**, 779-837(2014).

[2] C. Giannetti et al. Ultrafast optical spectroscopy of strongly correlated materials and high-temperature superconductors: a non-equilibrium approach. *Advances in Physics*, **65**, 58-238(2016).

[3] D. N. Basov, R. D. Averitt, & D. Hsieh, Towards properties on demand in quantum materials. *Nat. Matter.* 16, 1077-1088 (2017).

[4] T. F. Nova et al. An effective magnetic field from optically driven phonons, *Nat Phys.* **13**, 132(2017).

[5] M. Cammarata et al. Charge transfer driven by ultrafast spin transition in a CoFe Prussian blue analogue. *Nat. Chem.* 13, 10-14 (2021).

[6] S. Koshihara et al. Challenges in developing a new class of photoinduced phase transition (PIPT): a step from classical to quantum PIPT. *Phys. Rep.* **942**, 1-61(2022).

[7] A. Kitaev, Anyons in an exactly solved model and beyond. *Ann. Phys.* **321**, 2(2006).

[8] G. Jackeli, & G. Khaliullin, Mott insulators in the strong spin-orbit coupling limit: from Heisenberg to a quantum compass and Kitaev models. *Phys. Rev. Lett.* **102**, 017205(2009).

[9] S. M. Winter et al. Models and materials for generalized Kitaev magnetism. *J. Phys.: Condens. Matter* **29**, 493002(2017).

[10] H. Takagi et al. Concept and realization of Kitaev quantum spin liquid.

*Nat. Rev. Phys.* **1**, 264-280(2019).

[11] A. Kirrily, A. V. Kimel, & T. Rasing, Ultrafast optical manipulation of magnetic order. *Rev. Mod. Phys.* **82**, 2731-2784(2010).

[12] T. Kampfrath, K. Tanaka, K. A. Nelson, Resonant and nonresonant control over matter and light by intense terahertz transients, *Nat. Photon.* **7**. 680-690(2013).

[13] P. Němec, M. Fiebig, T. Kampfrath, & A. V. Kimel, Antiferromagnetic opto-spintronics. *Nat. Phys.* **14**, 229-241(2018).

[14] T. Satoh et al. Spin oscillation in antiferromagnetic NiO triggered by circularly polarized light *Phys. Rev. Lett.* **105**, 077402(2010).

[15] L. J. Sandilands et al. Scattering continuum and possible fractionalized excitations in α-$RuCl_3$. *Phys. Rev. Lett.* **114**, 147201(2015).

[16] A. Banerjee et al. Neutron scattering in the proximate quantum spin liquid α-$RuCl_3$. *Science* **356**, 1055-1059(2017).

[17] A. Little et al. Antiferromagnetic resonance and terahertz continuum in α-$RuCl_3$. *Phys. Rev. Lett.* **119**, 227201(2017).

[18] J. Nasu et al. Fermionic response from fractionalization in an insulating two-dimensional magnet. *Nat. Phys.* **12**, 912-916(2016).

[19] Y. Kasahara et al. Majorana quantization and half-integer thermal quantum Hall effect in a Kitaev spin liquid. *Nature* **559**, 227-232(2018).

[20] K. W. Plumb et al. α-$RuCl_3$ : A spin-orbit assisted Mott insulator on a honeycomb lattice. *Phys. Rev.* B**90**, 041112(2014).

[21] See Supplemental Material at http://link.aps.org/supplemental/ **.****/PhysRevResearch.*.L****** for (1) Variation of $T_N$ depending on

crystalline polymorph, (2) Dependence of $\Delta\theta$ on ellipticity of pump pulse, (3) Temperature dependence of $\Delta\theta$ above 20 K, (4) Helicity-independent slow component of $\Delta\theta$, (5) $\Delta\theta$ measurement under excitation of 1.55 eV (across the Mott gap), (6) Photo-carrier generation induced by excitation below Mott gap, (7) Calculated optical conductivity spectra for different values of SOI (Synergy of on-site Coulomb interaction and SOI), (8) Analysis of time evolutions of $\Delta T/T$ and $\Delta R/R$ by multi-exponential function, (9) Ultrafast carrier dynamics captured by 6-fs pulse, (10) Details of theoretical calculations: Model and parameters. (11) Details of theoretical analysis: High-frequency expansion in Floquet theory.

[22] J. A. Sears et al. Magnetic order in α-RuCl$_3$: A honeycomb-lattice quantum magnet with strong spin-orbit coupling. *Phys. Rev.* B**91**, 144420(2015).

[23] Y. Kubota et al. Successive magnetic phase transitions in α-RuCl$_3$: XY-like frustrated magnet on the honeycomb lattice. *Phys. Rev.* B**91**, 094422(2015).

[24] R. D. Johnson et al. Monoclinic crystal structure of α-RuCl$_3$ and the zigzag antiferromagnetic ground state. *Phys. Rev.* B**92**, 235119 (2015).

[25] H. B. Cao et al. Low-temperature crystal and magnetic structure of α-RuCl$_3$. *Phys. Rev.* B**93**, 134423(2016).

[26] S.-Y. Park et al. Emergence of the isotropic Kitaev honeycomb lattice with two-dimensional Ising universality in α-RuCl$_3$. *arXiv*:1609.05690v1 (19 Sep 2016).

[27] A. Glamazada et al. Relation between Kitaev magnetism and structure in α-RuCl$_3$. *Phys. Rev.* B**95**, 174429 (2017).

[28] Y. Nagai et al. Two-step gap opening across the quantum critical point in the Kitaev honeycomb magnet α-RuCl$_3$. *Phys. Rev.* B**101**, 020414(R)(2020).

[29] J. A. Sears et al. Phase diagram of α-RuCl$_3$ in an in-plane magnetic field. *Phys. Rev.* B**95**, 180411(R) (2017).

[30] L. Wu et al. Field evolution of magnons in α-RuCl$_3$ by high-resolution polarized terahertz spectroscopy. *Phys. Rev.* B**98**, 094425(2018).

[31] L, J. Sandilands et al. Spin-orbit excitations and electronic structure of the putative Kitaev magnet α-RuCl$_3$. *Phys. Rev.* B**93**, 075144(2016).

[32] P. Warzanowski et al. Multiple spin-orbit excitons and the electronic structure of α-RuCl$_3$. *Phys. Rev. Res.* **2**, 042007(R) (2020).

[33] J-H, Lee et al. Multiple spin-orbit excitons in α-RuCl$_3$ from bulk to automically thin layers. npj Quantum Mater. 6 43(2021).

[34] A. Loidl, P. Lunkenhaimer, & V. Tsurkan On the proximate Kitaev quantum-spin liquid α-RuCl$_3$: thermodynamics, excitations and continua. *J. Phys.: Condens. Matter* 33, 443004(2021).

[35] Y. Hasegawa et al. Two-phonon absorption spectra in the layered honeycomb compound α-RuCl$_3$. *J. Phys. Soc. Jpn.* **86**, 123709(2017).

[36] Y. Kawakami et al. Nonlinear charge oscillation driven by a single-cycle light field in an organic superconductor. *Nat. Photon.* **12**, 474-478(2018).

[37] Y. Kawakami et al., Petahertz non-linear current in a centrosymmetric organic superconductor. Nat. Commun. 11. 4138(2020).

[38] Y. P. Svirko, & N. I. Zheludev *Polarization of light in nonlinear optics.* John Willy & Sons, 1998.

[39] R. V. Mikhaylovskiy, E. Hendry, & V. V. Kruglyak, Ultrafast inverse Faraday effect in a paramagnetic terbium gallium garnet crystal. *Phys. Rev.* B**86**, 100405(R)(2012).

[40] R. B. Versteeg et al. Nonequilibrium quasistationary spin disordered state in α-RuCl$_3$, *Phys. Rev.* B105, 224428(2022).

[41] C. S. Wang, & J. Callaway, Band structure of nickel: Spin-orbit coupling, the Fermi surface, and the optical conductivity, *Phys. Rev.* B9, 4897(1974).

[42] D. Nevola et al. Timescales of excited state relaxation in α-RuCl$_3$. observed by time -resolved two-photon spectroscopy. *Phys. Rev.* B103, 245105(2021).

[43] J. G. Rau, E. K. -H. Lee, & H-Y. Kee, Generic spin model for the honeycomb iridates beyond the Kitaev limit. *Phys. Rev. Lett.* 112, 077204(2014).

[44] H.-S. Kim, & H. -Y. Kee Crystal structure and magnetism in α-RuCl$_3$: An ab initio study. *Phys. Rev.* B93, 155143 (2016).

[45] S. M. Winter, Y. Li, H. O. Jeschke, & R. Valenti Challenges in design of Kitaev materials. Magnetic interactions from competing energy scales. *Phys. Rev.* B93, 214431(2016).

[46] Y. Kawakami et al. Early-Stage Dynamics of Light-Matter Interaction Leading to the Insulator-to-Metal Transition in a Charge Ordered Organic Crysta. *Phys. Rev. Lett.* 105, 246402 (2010).

[47] T. Ishikawa et. al. Optical freezing of charge motion in an organic conductor. *Nat. Commun.* 5, 5528 (2014)

[48] K. Yonemitsu, S. Miyashita, & N. Maeshima Photoexcitation-energy-dependent transition pathways from a dimer Mott insulator to a metal. *J. Phys. Soc. Jpn.* 80, 084710 (2011).

[49] K. Yonemitsu, & N. Maeshima Coupling-dependent rate of energy

transfer from photoexcited Mott insulators to lattice vibrations, *Phys. Rev.* B**79**, 125118 (2009).

[50] A. Eckardt, & E. Anisimovas, High-frequency approximation for periodically driven quantum systems from a Floquet-space perspective, *N. J. Phys.* **17**, 093039 (2015).

**Figure Legends**

**Fig. 1 a** Honeycomb lattice of $\alpha$-RuCl$_3$ and hybridization between $t_{2g}$ orbitals of Ru and ligand $p_z$ orbital of Cl which causes hopping $t$ between different $t_{2g}$ orbitals (such as $d_{xz} - d_{yz}$). **b** Imaginary part of the permittivity spectrum in mid- and near- infrared regions [31]. The low-energy part below 0.9 eV (magnified 25 times) is attributed to SOC in-gap states [31-33]. The experimental configuration of the opto-magneto measurement (polarization rotation induced by a circularly polarized pump pulse) and the excitation from $J_{eff}$=3/2 to $J_{eff}$=1/2 states [32] are schematically shown.

**Fig. 2 a** Time evolutions of $\Delta\theta$ (upper panel) and transmittance ($\Delta T/T$) (lower panel) of $\alpha$-RuCl$_3$ ($E_{pu}$=0.89 eV, $I_{ex}$=4 mJ/cm$^2$, $E_{pr}$=0.62 eV, ∥ab plane, $\sigma^+$: red curves, $\sigma^-$: blue curves) at 4 K (dashed curves) and 17 K (solid curves). **b** Temperature dependences of $\Delta\theta$ at $t_d$= 0 ps (upper panel) and at $t_d$= 10 ps (lower panel). **c, d** Opto-magneto spectra (17 K) of $\Delta\theta$ (**c**) and $\Delta\eta$ (**d**) for (i) $E_{pu}$=0.30 eV, (ii) $E_{pu}$=0.62 eV, and (iii) $E_{pu}$=0.89 eV. The upper panel of **c** shows the imaginary permittivity spectrum [31] (x 25 for < 0.9 eV). The maximum values of $\Delta\theta$ for $E_{pu}$=0.30-0.89 eV are shown by the red dots. A schematic illustration of the opto-magneto 2-step process is shown in the upper panel of **d**.

**Fig. 3 a** Imaginary part of the permittivity spectrum at 10 K [31] around the Mott gap. **b, c** $\Delta R/R$ spectra for $E_{pu}$=0.62 eV (**b**) and $E_{pu}$=0.89 eV (**c**). $I_{ex}$=1 mJ/cm$^2$, 17 K, $t_d$=0 ps (red dots), 0.5 ps (orange circles), 1 ps (green circles).

The insets of (b) and (c) show the $I_{ex}$ dependence of $\Delta R/R$ at $E_{pr}$=1.2 eV. The circles with the dashed curve in (c) show $\Delta R/R$ for $E_{eu}$=1.55 eV (4 mJ/cm$^2$ ,x 0.17, 4 K, $t_d$=0 ps ), $E_{pr}$ =1.2 eV. **d** Time profiles of $\Delta R/R$ ($E_{pr}$= 1.2 eV) for $E_{pu}$=0.62 eV (blue circles) and $E_{pu}$=0.89 eV (red circles x0.6). **e** Time profile of $\Delta R/R$ measured by using a 6 fs near-infrared pulse ($E_{pu}$=0.55-0.95 eV, $E_{pr}$=0.60 eV (with polarization perpendicular to that of the pump pulse), $I_{ex}$=1.0 mJ/cm$^2$, 17 K). The red curve shows exponential decay with a time constant of 0.17 ps. The oscillating component is shown in the inset (green circles: 17 K, blue circles 300 K, red curve: oscillation with a period of 36 fs (Supplementary 9 [21])).

**Fig. 4 a** Honeycomb lattice structure of α-RuCl$_3$ used in the calculation. It has three unit cells, each of which consists of two sites A and B. Periodic boundary conditions are used to maintain the threefold symmetry. X1, Y1, and Z1 connect nearest-neighbor sites, and X2, Y2, and Z2 connect next-nearest-neighbor sites. **b** Calculated optical conductivity spectra for the ground state (red and blue curves are used for ||a, ||b). The spectra below 0.9 eV are magnified by 15 times. **c** Calculated time profiles of photoinduced magnetization in the direction of ⊥ab. An electric field amplitude times intersite distance F (V) of 0.2 (an electric field amplitude of 5.8 MV/cm), frequencies ℏω (eV) of 0.3, 0.6, and a pulse width of 96 fs are used.

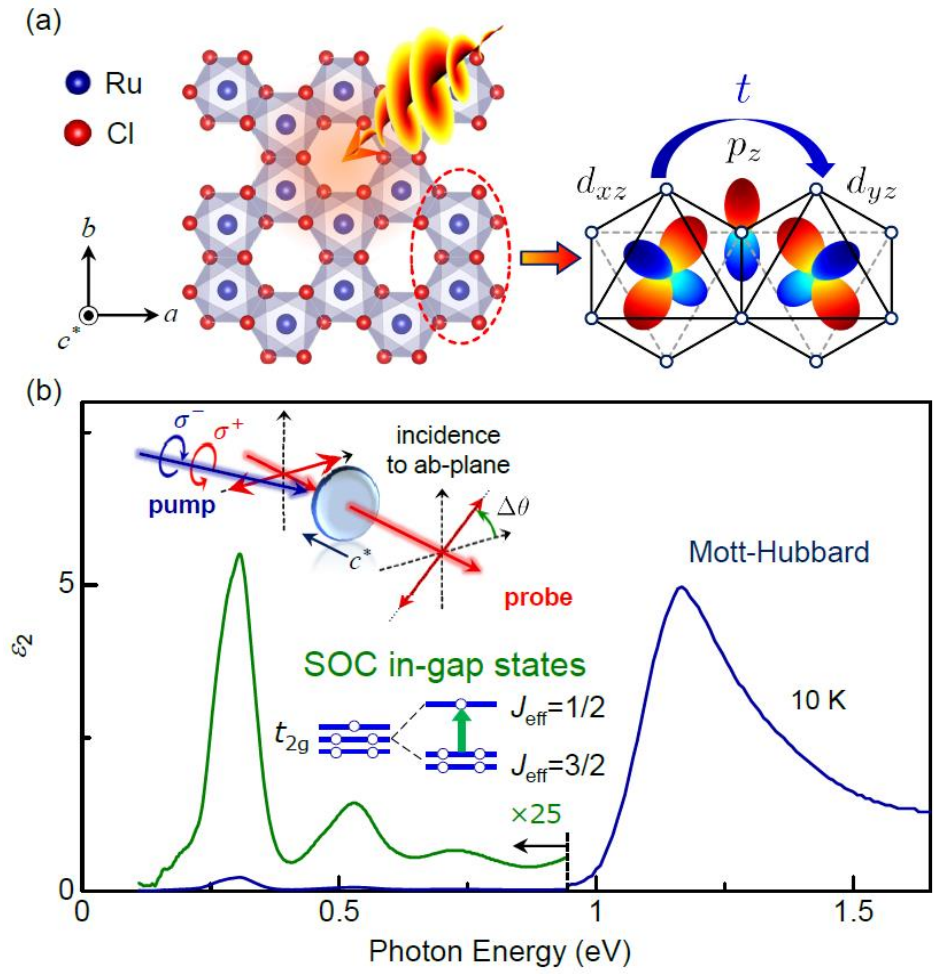

**Figure 1**

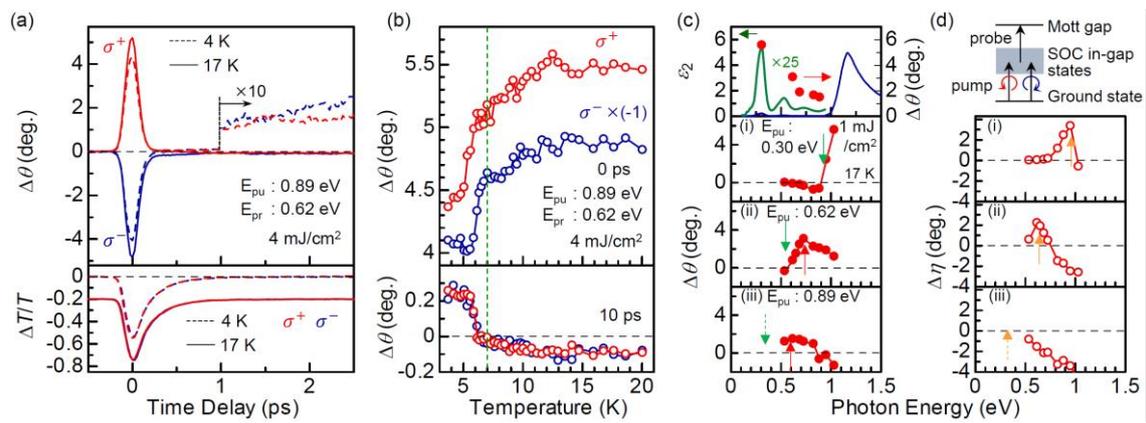

**Figure 2**

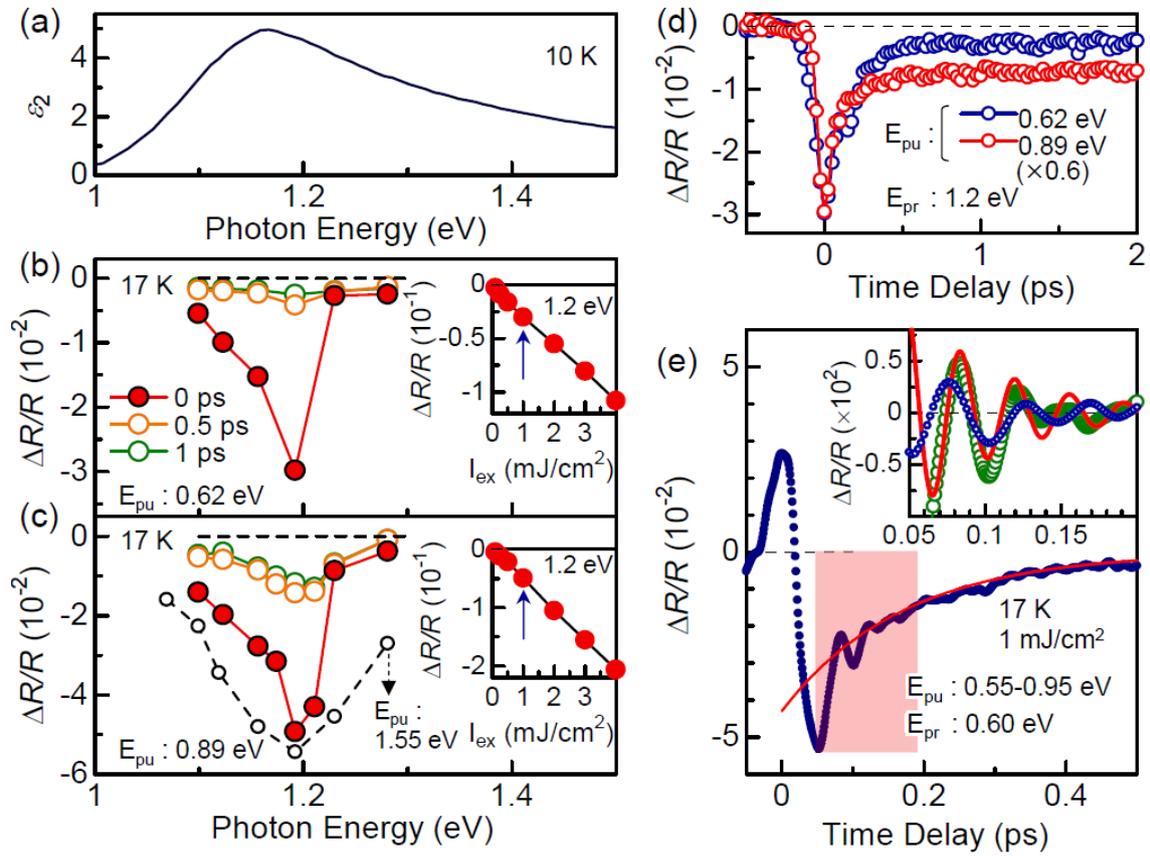

Figure 3

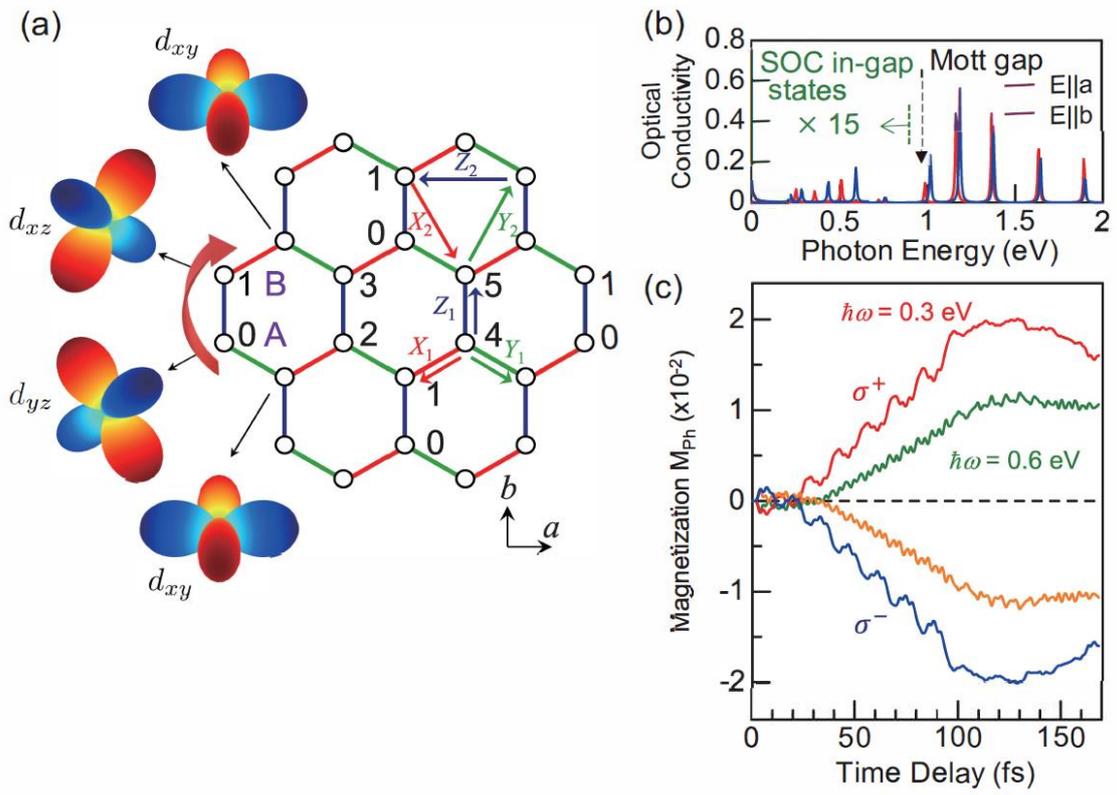

Figure 4

Supplementary information for

# Light-induced magnetization driven by interorbital charge motion in a spin-orbit assisted Mott insulator α-RuCl₃

T. Amano[1], Y. Kawakami[1], H. Itoh[1], K. Konno[1], Y. Hasegawa[1], T. Aoyama[1],

Y. Imai[1], K. Ohgushi[1], Y. Takeuchi[1], Y. Wakabayashi[1], K. Goto[2],

Y. Nakamura[2], H. Kishida[2], K. Yonemitsu[3], and S. Iwai[1]*

[1]Department of Physics, Tohoku University, Sendai 980-8578, Japan

[2] Department of Applied Physics, Nagoya University, Nagoya 464-8603 Japan

[3]Department of Physics, Chuo University, Tokyo 112-8551, Japan

* s-iwai@tohoku.ac.jp

## Supplementary note 1
### Variation of $T_N$ depending on crystalline polymorph

The magnetic transition into the zigzag AF order in α-RuCl₃ has been controversial in the sense that the stacking sequence along the c-axis has not been established at low temperature because various polytypes with different stacking patterns are almost energetically degenerate [22-25]. A recent report shows the low temperature structure of $R\bar{3}$ [23, 26-28] with the magnetic transition temperature $T_N$=7 K. Some crystals have a portion that has $C2/m$ space group; thus, the transition temperature ranges $T_N$ =7 K to 14 K.

We have checked that the volume fraction of $R\bar{3}$ is ca. 3/4 and the magnetic transition occurs dominantly at 7 K in our samples by measuring the XRD and magnetic susceptibility. We also check the steady-state polarization rotation angle $\theta$ to confirm the $T_N$ value. Figure S1(a) shows the relative angle ($\theta_R(T) = \theta(T) - \theta(20\ K)$) of a transmitted 0.62 eV pulse for several samples (#1, #2, #3) as a function of temperature $T$. The finite $\theta_R$ in the AF phase is

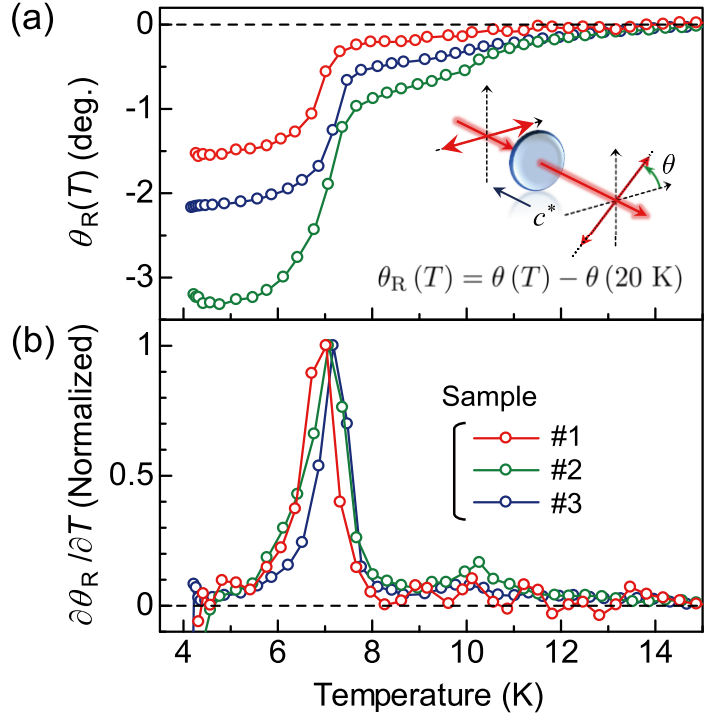

**Fig. S1** (a) Temperature dependence of polarization rotation angles ($\theta_R = \theta(T) - \theta(20\,K)$) at 0.62 eV for different α-RuCl$_3$ samples (#1(red circles), #2(green circles), and #3(blue circles)). (b) $\partial\theta_R/\partial T$ indicating abrupt changes in $\theta_R$ for the respective samples at 7 K.

attributed to the Cotton-Mouton effect induced by the AF vector ($\boldsymbol{L} = \boldsymbol{M}_1 - \boldsymbol{M}_2$) where $\boldsymbol{M}_1$ and $\boldsymbol{M}_2$ represent the magnetic moments for the respective magnetic sublattices. Considering that three-types of possible magnetic domains have been discussed in this compound [29, 30] regardless of the above-mentioned issue on the space group, the observed $\theta_R$ reflects the unbalanced distribution of the respective domains within the probing spot with a diameter of 100 microns. In fact, $\theta_R$ sensitively depends on the sample.

As shown in Fig. S1(a), the $\theta_R$ shows an abrupt change near $T_N$ with increasing temperature where $\partial\theta_R/\partial T$ indicates a maximum at 7 K (Fig. S1(b)). Then, the small remaining component gradually decreases in magnitude toward zero until $T=14$ K. Thus, the magnetic transition occurs dominantly at $T_N = 7$ K in our samples. We have actually chosen the crystal in which the $T_N=14$ K component is the smallest.

Considering that the anomaly in $\theta_R$ at 7 K is actually confirmed in the steady state, the shift of the temperature where the abrupt change of $\Delta\theta$ occurs in Fig. 2(b) (which is approximately 1 K lower than $T_N$=7 K) is attributed to the heat accumulation by the pump light in 1 kHz operation.

The optical anisotropy in the ab-plane just reflects the sample quality, i.e., the volume fraction of $R\bar{3}$, and it is small and non-essential for the present study of the photoinduced magnetization and the photocarrier dynamics. Thus, we do not discuss the dependence of $\Delta\theta$ on the polarization of the probe pulse following the previous reports [31-35, 40].

Supplementary note 2
Dependence of $\Delta\theta$ on ellipticity of pump pulse

The magnetization component perpendicular to the ab-plane is caused by the circularly polarized light in Fig. 2(a). It is phenomenologically called the inverse Faraday effect (IFE). In the measurement of the polarization rotation induced by the IFE, the pump light should be strictly circularly polarized inside the material. If the polarization of the pump light becomes elliptic during the propagation in the material, the optical Kerr effect (OKE) also gives a polarization rotation [38, 39]. To confirm the IFE, we need to check that the OKE is absent in the figures shown in the main text. For this purpose, we measure the dependence of $\Delta\theta$ on the ellipticity of the pump light.

The ellipticity of the pump pulse is changed by using a λ/4 plate (SAQWP05M-1700 for the range 600-2700 nm (Thorlabs, Inc.) and a zero order waveplate for 4 µm (Edmund Optics Inc.)). The linearly polarized pump pulse enters into the λ/4 plate with an angle $\delta$ between the polarization direction of the incident pump pulse and the optical axis (the fast axis) of the λ/4 plate. The dots in Fig. S2(a) show $\Delta\theta$ (17 K (spin-liquid phase), $E_{pu}$=0.89 eV, $E_{pr}$=0.62 eV, $I_{ex}$=1.0 mJ/cm) as a function of $\delta$. Here, the polarization of the probe pulse is parallel to that of the original pump pulse before entering the λ/4 plate. In principle, an outgoing light from the λ/4 plate should be circularly polarized for $\delta$ =45 and 135 degrees, and it should be linearly

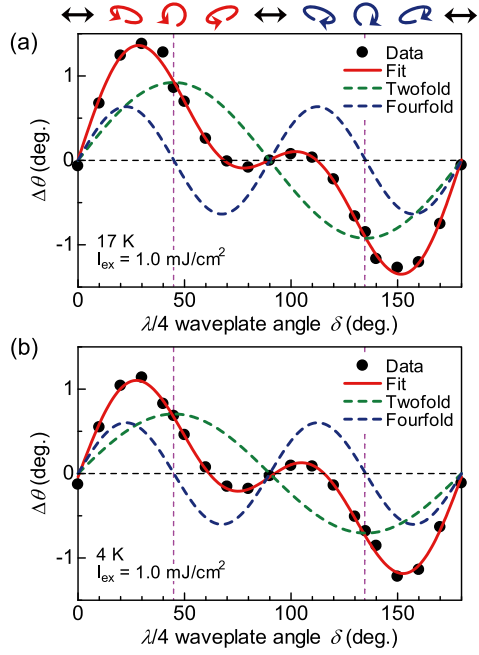

**Fig. S2** Dependence of $\Delta\theta$ on $\delta$ (the angle between the polarization direction of the original linearly polarized pump pulse and the optical axis of the λ/4 plate for changing ellipticity). The polarization of the probe pulse is parallel to that of the original pump pulse. The measurements are made in the spin-liquid phase (**a** 17 K) and in the AF phase (**b** 4 K) ($E_{pu}$=0.89 eV, $E_{pr}$=0.62 eV, $I_{ex}$=1.0 mJ/cm²). The twofold and fourfold natures are indicated by the green and the blue dashed curves, respectively. The red curves show the linear combination of them.

polarized for $\delta$ =0 and 90 degrees as schematically shown in Fig. S2.

As shown in Fig. S2(a), the ellipticity dependence of $\Delta\theta$ is reproduced by the linear combination of the twofold (green dashed curve) and the fourfold (blue dashed curve) functions as $\Delta\theta = a\sin(2\delta) + b\sin(4\delta)$ ($a$=0.92, $b$=0.64) where $a$ and $b$ represent the coefficients of the IFE (twofold) and the OKE (fourfold). The contribution from the IFE is actually dominant for $\delta$=45 and 135 degrees (as shown by the dashed magenta lines), while the OKE shows maxima for the elliptic polarization $\delta$= 22.5, 67.5, 112.5, 157.5 degrees. This fact clearly confirms that the contribution from the OKE is not included in $\Delta\theta$ shown in Fig. 2(a) (for $\delta$=45 degrees).

As shown in Fig. S2(b), the similar result is observed also in the AF phase (4 K). It is noteworthy that the coefficient a (4 K) =0.71 is smaller than that of 0.92 at 17 K, although b (4K) =0.60 is approximately equal to 0.64 at 17 K. Such temperature-sensitive nature of the IFE and temperature-insensitive nature of the OKE are quite reasonable.

Supplementary note 3
Temperature dependence of $\Delta\theta$ above 20 K

In Fig. S3, the temperature dependences of $\Delta\theta$ are shown above 20 K ($E_{pu}$=0.89 eV, 4 mJ/cm, $E_{pr}$=0.62 eV) for $\sigma^+$(red circles) and $\sigma^-$ (blue circles). In contrast to the temperature-sensitive nature around $T_N$=7 K, $\Delta\theta$ is insensitive to the temperature above 20 K.

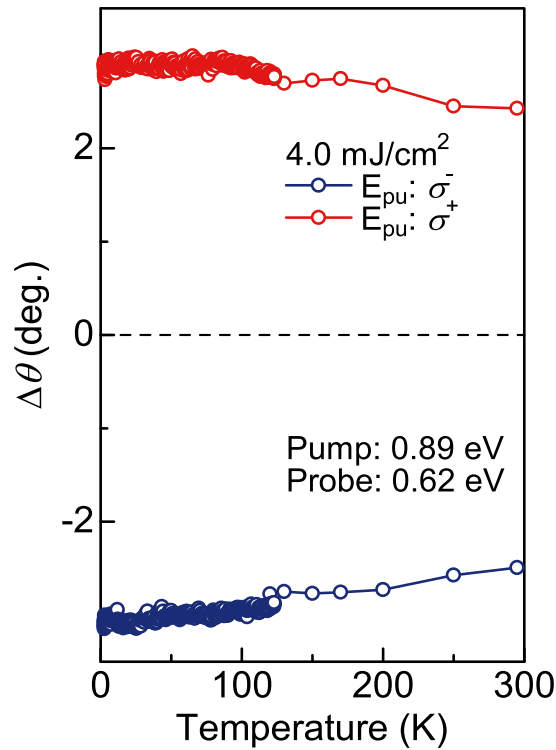

**Fig. S3** Temperature dependences of $\Delta\theta$ above 20 K ($E_{pu}$=0.89 eV, 4 mJ/cm, $E_{pr}$=0.62 eV) for $\sigma^+$(red circles) and $\sigma^-$ (blue circles).

Supplementary note 4

Helicity-independent slow component of $\Delta\theta$

The helicity-independent growing-up component on the time scale of 10 ps only for the AF phase at $T < T_N$ is attributed to the melting of the AF order. As described in Supplementary 1, a finite $\theta_R$ (the relative rotation angle of the probe pulse in the steady state) reflects the unbalanced distribution of the three types of possible magnetic domains within the probing spot with a diameter of 100 microns at $T < T_N$ [29, 30]. The excitation with 4 mJ/cm² causes the transient lattice temperature rise (ca. 40 K) (on the basis of the specific heat, the penetration depth, the reflection loss, and the unit cell volume). It easily melts the AF order. The time scale of the slow component (ca. 10 ps) is consistent with the energy scale of the magnon (ca. 1 meV=h/(4 ps)). This slow component of $\Delta\theta$ is observed also for the excitation across the Mott gap ($E_{pu}$=1.55 eV for both linear and circular polarizations), where the fast component due to the ultrafast magnetization is not detected. Recovery processes from this photo-induced melting of the AF state have been investigated [40].

Supplementary note 5

Δθ measurement under excitation of 1.55 eV (across the Mott gap)

In Figs. 2(a), 2(b), and 2(c), the excitation energies are 0.3, 0.62, 0.89 eV (below the Mott gap, circular polarization) and the probe energies are 0.5-1 eV. We also perform measurements of Δθ under the excitation of 1.55 eV (across the Mott gap).

As shown in Fig. S4, the time evolution of Δθ ($E_{pu}$=1.55 eV (1 mJ/cm$^2$, circular ($\sigma^+$:red curve, $\sigma^-$: blue curve) and linear (green curve) polarizations), $E_{pr}$=0.54 (a) and 1.03 eV(b)) shows Δθ~0.02 degrees at most for any polarization which is <1/50 in comparison with those of $E_{pu}$=0.89 eV and $E_{pr}$=0.62 eV (1 mJ/cm$^2$). Furthermore, the helicity-dependent component is very small (the difference between Δθ for $\sigma^+$ (red curve) and Δθ for $\sigma^-$ (blue curve) is 0.005 degrees) so that it is < 1/200. They are negligible in comparison with the < 100 fs component in our results.

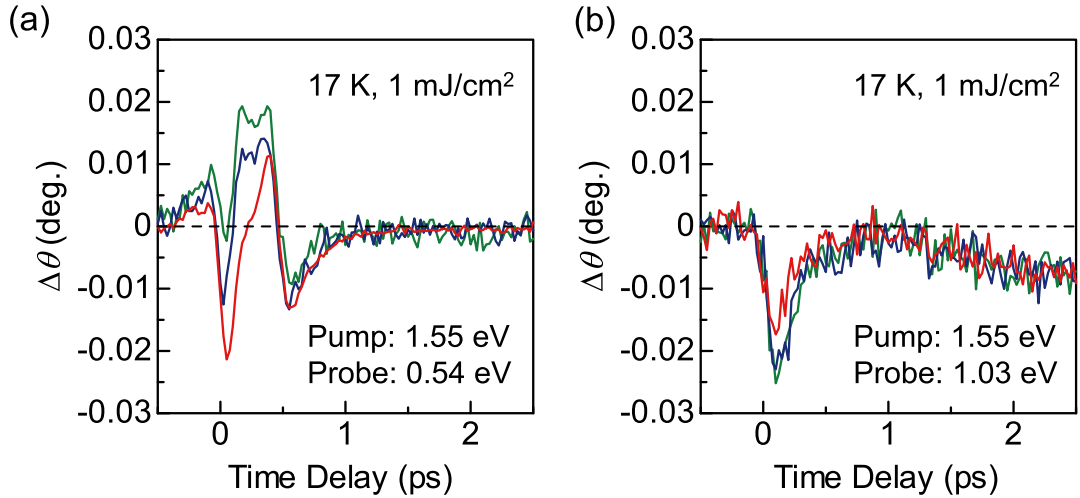

**Fig. S4** Time evolution of Δθ ($E_{pu}$=1.55 eV(1 mJ/cm$^2$, circular ($\sigma^+$:red curve, $\sigma^-$: blue curve) and linear (green curve) polarizations), $E_{pr}$=0.54 (a) and 1.03 eV(b))

Supplementary note 6

Photo-carrier generation induced by excitation below Mott gap

In ref. [42], a two-photon photoemission spectroscopy and a transient reflectivity measurement clearly show the generation of carriers (doublon - holon pairs) for the excitation above the Mott gap. We actually confirm that the bleaching around the Mott transition peak for $E_{pu}$=1.55 eV (> $E_{gap}$, circularly polarized, 4 K, circles with dashed curve in Fig. 3(c)) is consistent with the results in ref. [42], although the excitation is made by a linearly polarized pump pulse at RT in ref. [42]. As shown in the main text, the similar results are obtained for the in-gap excitations as shown in Figs. 3(b) ($E_{pu}$=0.62 eV) and 3(c) ($E_{pu}$=0.89 eV). They are not attributable to the two-photon (2x $E_{pu}$) absorption as the origin of the bleaching because of their linear dependence on $I_{ex}$ as shown in the insets of Figs. 3(b) and 3(c). Considering the nature of the SOI assisted Mott insulator, the results in Figs. 3(b) and 3(c) indicate that carriers are generated even by the in-gap excitation [Supplementary 7].

Supplementary note 7

Calculated optical conductivity spectra for different values of SOI (Synergy of on-site Coulomb interaction and SOI)

The Mott insulating state of this compound is realized by the synergy of the on-site Coulomb interaction $U$ and the SOI [20]. This fact is confirmed within the theoretical framework of this study [Supplementary 10] as described below. Figures S5(a) and S5(b) show calculated optical conductivity spectra for polarizations ||**a** (Fig. S5(a)) and ||**b** (Fig. S5(b)) for different values of the SOI (λ=0.15 (black circles) (which is equal to the value in ref. [45].), 0.10 (blue curves), 0.06 (green curves), and 0.02 eV (red curves)). The other model parameters including $U$=3.0 eV, $J_H$=0.6 eV, and $U'$=1.8 eV are the same as those in ref. [45] and those used in Supplementary 10 for Fig. 4 ($U$: intra-orbital Coulomb interaction, $J_H$ :Hund's coupling, $U' = U – 2J_H$). We notice the Mott gap of ca. 1 eV for λ=0.15 eV. With reducing λ, the Mott gap or the remnant of the Mott gap decreases as shown by the bule, green, and red arrows. In addition, the oscillator strengths of the in-gap states grow with decreasing λ suggesting the tendency toward the collapse of the Mott gap, while they are very small for λ=0.15 eV. Thus, the SOI is clearly demonstrated to support the insulating gap for α-RuCl$_3$ in Fig. S5. Carriers are then naturally generated by the excitation of the SOC in-gap states.

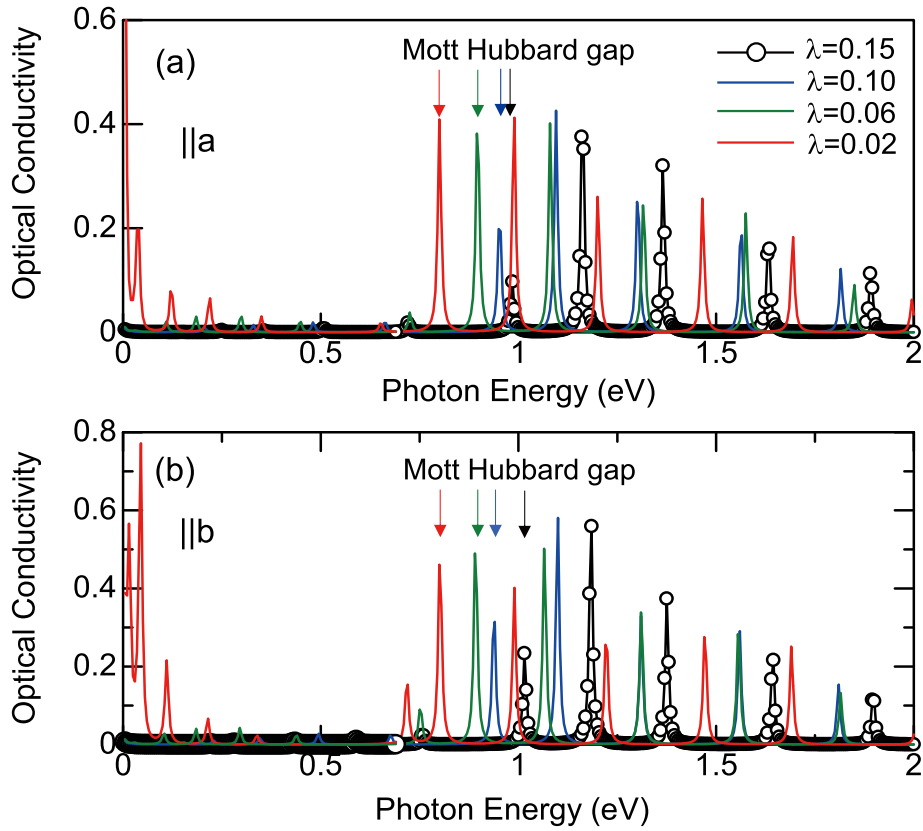

**Fig. S5** Calculated optical conductivity spectra for polarizations ||a (**a**) and ||b (**b**) for λ=0.15 (black circles) (which is equal to the value in ref. [45].), 0.10 (blue curves), 0.06 (green curves), and 0.02 eV (red curves). The other model parameters are the same as those in ref. [45] and those used in Supplementary 10 for Fig. 4.

Supplementary note 8

Analysis of time evolutions of $\Delta T/T$ and $\Delta R/R$ by multi-exponential function

We have analyzed the time profile of $\Delta T/T$ (Fig. 2(a) lower panel)) and those of $\Delta R/R$ (Fig. 3(d)) by using the following equations.

$$\frac{\Delta R(t)}{R} \text{ or } \frac{\Delta T(t)}{T} = A_0 G(t) + \int_{-\infty}^{+\infty} K(t)G(t-t')dt',$$

$$K(t) = \sum_{i=1}^{n} A_i [1-\exp(-t/\tau_{ri})] \exp\left(-\frac{t}{\tau_{di}}\right),$$

$$G(t) = \exp[-4ln2\, t^2/\{(Pw\,(fs))^2\}],$$

where G(t) is a gaussian with a width of the instantaneous response ($Pw$=150 fs for $\Delta T/T$, $Pw$=100 fs ($E_{pu}$=0.62 eV), 70 fs ($E_{pu}$=0.89 eV) for $\Delta R/R$). $A_i$, $\tau_{ri}$, $\tau_{di}$ represent the amplitude, rise and decay time constants of the $i$ th component, respectively (n=1 for $\Delta T/T$, n=2 for $\Delta R/R$). $A_0$ is the amplitude of the instantaneous (as fast as 100 fs) response.

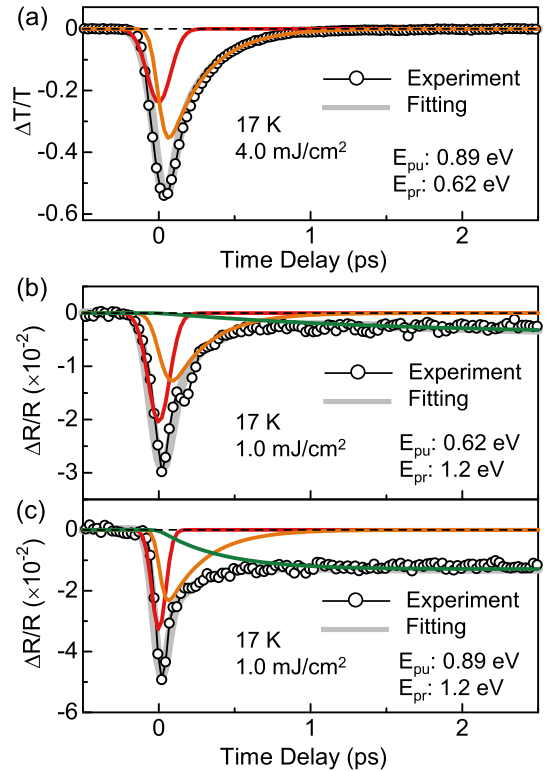

**Fig. S6** Analysis of time evolutions by multi-exponential functions for $\Delta T/T$ (a) and $\Delta R/R$ (b)(c) (See Supplementary 8).

Figure S6(a) shows that the time profile of $\Delta T/T$, (Fig. 2(a), lower panel) is reproduced by $A_0$=-0.29 (red line) for the instantaneous response and $A_1$=-0.5, $\tau_{r1}$=0, $\tau_{d1}$=0.25 ps (orange line). Thus, the $\Delta T/T$ is characterized by the decay time constant of 0.25 ps in addition to the instantaneous gaussian response. On the other hand, Figs. S6(b) and S6(c) show the time profiles of $\Delta R/R$ (Fig. 3(d)). They are reproduced by $A_0$=-0.034 (red line), $A_1$=-0.02, $\tau_{r1}$=0 ps, $\tau_{d1}$=0.25 ps (orange line), $A_2$=-0.004, $\tau_{r2}$=1.6 ps, $\tau_{d2}$= > 1ns (green line) for $E_{pu}$ =0.62 eV, and $A_0$=-0.05 (red line), $A_1$=-0.032, $\tau_{r1}$=0 ps, $\tau_{d1}$=0.25 ps (orange line), $A_2$=-0.013, $\tau_{r2}$=0.4 ps, $\tau_{d2}$=> 320 ps (green line) for $E_{pu}$ =0.89 eV.

According to the above analysis, the time constants for the time profiles of $\Delta R/R$ ($E_{pu}$=0.89, 0.62 eV, $E_{pr}$=1.2 eV) (Fig. 3(d)) are comparable with those of $\Delta T/T$, indicating that a large $\Delta T/T$<0 reflects the photo-carrier generation.

The instantaneous gaussian responses in $\Delta T/T$ and $\Delta R/R$ correspond to the response as fast as 100 fs in $\Delta\theta$ in the upper panel of Fig. 2(a). The decay components $d$1 (0.25 ps in $\Delta R/R$ and $\Delta T/T$) and $d$2 (1 ns for $E_{pu}$=0.62 eV and 320 ps for $E_{pu}$= 0.89 eV in $\Delta R/R$) are attributed to the charge relaxation processes mediated by phonons and/or spins.

Supplementary note 9
### Ultrafast carrier dynamics captured by 6-fs pulse

Figure 3(e) shows the time evolution of $\Delta R/R$ measured at 0.60 eV (under the excitation of a linearly polarized 6 fs pump pulse covering the spectral range of 0.55-0.95 eV with $I_{ex}$=1 mJ/cm$^2$). In this measurement, each of the pump and probe lights is resonant to SOC in-gap states (or equivalently an excitation from $J_{eff}$=3/2 to $J_{eff}$=1/2 states). Therefore, it is natural to consider that we detect the modulation of the SOC in-gap states in this experimental configuration. The time profile is reproduced by the single exponential function with a time constant of 0.17 ps (red line), except for a time region around $t_d$=0 which is affected by a coherent polarization effect such as a perturbed free induction decay. The time constant of 0.17 ps is comparable with the time constant (0.25 ps) of $\Delta R/R$ ($E_{pu}$=0.62 and 0.89 eV, Fig. 3(d)) and $\Delta T/T$ ($E_{pu}$=0.89 eV, $E_{pr}$=0.62 eV, Fig. 2(a)). They are attributed to the relaxation of excited charges mediated by phonons and/or spins. Furthermore, the oscillating component (inset of Fig. 3(e)) is obtained by subtracting the exponential function from the observed curve. The oscillation after 50 fs is reproduced by $-0.024 \sin \frac{2\pi}{36\,fs}(t - 3\,fs)\exp\left(-\frac{t}{60 fs}\right)$ as shown by the red curve in the inset. Although, the phonon energies are restricted to the range below 40 meV in α-RuCl$_3$ [15-17, 34], higher orders of phonons such as an overtone should also be considered as an origin of the 36 fs (=h/(0.11 eV)) oscillation. However, the amplitudes of higher orders are usually much smaller than those of the fundamentals unless they are resonant to an electronic transition. In fact, there is no electronic transition around 0.11 eV [32-35] at low temperatures. Therefore, we can rule out the possibility of higher order phonon processes.

The oscillation reflects a coherent intersite charge motion, as can be understood by the fact the oscillating period (36 fs=h/(0.11 eV)) corresponds to the energies of the intersite hopping (0.1-0.2eV). Such coherent charge motions in correlated systems are often realized by the simultaneous application of a light-field to many electrons that are correlated by the Coulomb interactions [46, 47]. In strongly correlated systems, the electronic oscillation often loses its coherence within an ultrafast time scale through the strong electron-electron and electron-phonon interactions. This is consistent with the broad spectrum in the steady state of these compounds. In fact, the bandwidth of the $\varepsilon_2$ spectrum shown in Fig. 1(a) (ca. 100 meV) is consistent with the the coherence time (60 fs= h/ 70 meV) of the electronic oscillation. Thus, the short-lived oscillation is reasonable and essential in this compound.

Note that the optical transitions between $J_{eff}$=1/2 and $J_{eff}$=3/2 involving the in-gap states are realized by the intersite charge hopping. Therefore, it is quite reasonable to conclude that the intersite charge hopping and the SOC interplay. In fact, the energies of the intersite hopping and the SOI are comparable: 0.1-0.2 eV.

Supplementary note 10

Details of theoretical calculations: Model and parameters

We consider a Hubbard model consisting of $d_{yz}$, $d_{xz}$, and $d_{xy}$ orbitals on each site of the honeycomb lattice [43-45] using the model parameters employed in ref. [45]. The hopping term is limited to the nearest neighbors because the 6-site system is treated by the exact diagonalization method. Numerically we consider a minimum-size system that has three unit cells, each of which consists of two sites A and B, and use periodic boundary conditions shown in Fig. 4(a) (X1, Y1, and Z1 connect nearest-neighbor sites, and X2, Y2, and Z2 connect next-nearest-neighbor sites) to maintain the threefold symmetry.

With one hole per site, the three-orbital Hubbard model is written as,

$$H = H_{\text{hop}} + H_{\text{CF}} + H_{\text{SO}} + H_U,$$

which consists of the kinetic term, the crystal-field term, the spin-orbit coupling, and Coulomb interactions, respectively. With the use of

$$\vec{c}_i^\dagger = \left(c_{i,yz,\uparrow}^\dagger c_{i,yz,\downarrow}^\dagger c_{i,xz,\uparrow}^\dagger c_{i,xz,\downarrow}^\dagger c_{i,xy,\uparrow}^\dagger c_{i,xy,\downarrow}^\dagger\right),$$

where $c_{i,a,\sigma}^\dagger$ creates a hole in orbital $a$ with spin $\sigma$ at site $i$, the kinetic term is written as

$$H_{\text{hop}} = -\sum_{ij} \vec{c}_i^\dagger \{\mathbf{T}_{ij} \otimes \mathbb{I}_{2\times 2}\} \vec{c}_j.$$

Here, $\mathbb{I}_{2\times 2}$ is the $2 \times 2$ identity matrix and $\mathbf{T}_{ij}$ is the hopping matrix defined for each bond connecting nearest-neighbor sites $i$ and $j$. The latter is one of $\mathbf{T}_1^X$, $\mathbf{T}_1^Y$, and $\mathbf{T}_1^Z$ for the $X_1$, $Y_1$, and $Z_1$ bonds, respectively, shown in Fig. 4(a):

$$\mathbf{T}_1^X = \begin{pmatrix} t_3' & t_{4a}' & t_{4b}' \\ t_{4a}' & t_{1a}' & t_2' \\ t_{4b}' & t_2' & t_{1b}' \end{pmatrix}, \mathbf{T}_1^Y = \begin{pmatrix} t_{1a}' & t_{4a}' & t_2' \\ t_{4a}' & t_3' & t_{4b}' \\ t_2' & t_{4b}' & t_{1b}' \end{pmatrix}, \mathbf{T}_1^Z = \begin{pmatrix} t_1 & t_2 & t_4 \\ t_2 & t_1 & t_4 \\ t_4 & t_4 & t_3 \end{pmatrix}.$$

The values of the transfer integrals and the other parameters below are taken from ref. [45]: $t_1 = 0.0509$ eV, $t_{1a}' = 0.0449$ eV, $t_{1b}' = 0.0458$ eV, $t_2 = 0.1582$ eV, $t_2' = 0.1622$ eV, $t_3 = -0.1540$ eV, $t_3' = -0.1031$ eV, $t_4 = -0.0202$ eV, $t_{4a}' = -0.0151$ eV, and $t_{4b}' = -0.0109$ eV. The crystal-field term is written as

$$H_{\text{CF}} = -\sum_i \vec{c}_i^\dagger \{\mathbf{E}_i \otimes \mathbb{I}_{2\times 2}\} \vec{c}_i,$$

where $\mathbf{E}_i$ is the crystal-field tensor given by

$$\mathbf{E}_i = \begin{pmatrix} 0 & \Delta_1 & \Delta_2 \\ \Delta_1 & 0 & \Delta_2 \\ \Delta_2 & \Delta_2 & \Delta_3 \end{pmatrix},$$

with $\Delta_1 = -0.0198$ eV, $\Delta_2 = -0.0175$ eV, and $\Delta_3 = -0.0125$ eV. The spin-orbit coupling is written as

$$H_{\text{SO}} = \frac{\lambda}{2} \sum_i \vec{c}_i^\dagger \begin{pmatrix} 0 & -i\sigma_z & i\sigma_y \\ i\sigma_z & 0 & -i\sigma_x \\ -i\sigma_y & i\sigma_x & 0 \end{pmatrix} \vec{c}_i,$$

where $\sigma_x$, $\sigma_y$, and $\sigma_z$ are the Pauli matrices and $\lambda = 0.15$ eV. The Coulomb interactions are written as

$$H_U = U \sum_{i,a} n_{i,a,\uparrow} n_{i,a,\downarrow} + (U' - J_{\text{H}}) \sum_{i,a<b,\sigma} n_{i,a,\sigma} n_{i,b,\sigma} + U' \sum_{i,a\neq b} n_{i,a,\uparrow} n_{i,b,\downarrow}$$
$$- J_{\text{H}} \sum_{i,a\neq b} c_{i,a,\uparrow}^\dagger c_{i,a,\downarrow} c_{i,b,\downarrow}^\dagger c_{i,b,\uparrow} + J_{\text{H}} \sum_{i,a\neq b} c_{i,a,\uparrow}^\dagger c_{i,a,\downarrow}^\dagger c_{i,b,\downarrow} c_{i,b,\uparrow},$$

where $n_{i,a,\sigma} = c_{i,a,\sigma}^\dagger c_{i,a,\sigma}$, $U$ is the intraorbital Coulomb repulsion, $J_{\text{H}}$ is the Hund's coupling strength, and $U' = U - 2J_{\text{H}}$ is the interorbital repulsion with $U = 3.0$ eV and $J_{\text{H}} = 0.6$ eV.

To describe $J_{\text{eff}} = \frac{1}{2}$ pseudospin states, we introduce

$$p_{i,\uparrow}^\dagger = \frac{1}{\sqrt{3}} \left( -c_{i,xy,\uparrow}^\dagger - i c_{i,xz,\downarrow}^\dagger - c_{i,yz,\downarrow}^\dagger \right),$$

$$p_{i,\downarrow}^\dagger = \frac{1}{\sqrt{3}} \left( c_{i,xy,\downarrow}^\dagger + i c_{i,xz,\uparrow}^\dagger - c_{i,yz,\uparrow}^\dagger \right),$$

to define the pseudospin densities

$$j_{i,x}^{(1/2)} = \frac{1}{2} \langle p_{i,\uparrow}^\dagger p_{i,\downarrow} + p_{i,\downarrow}^\dagger p_{i,\uparrow} \rangle,$$

$$j_{i,y}^{(1/2)} = \frac{1}{2} \langle -i p_{i,\uparrow}^\dagger p_{i,\downarrow} + i p_{i,\downarrow}^\dagger p_{i,\uparrow} \rangle,$$

$$j_{i,z}^{(1/2)} = \frac{1}{2} \langle p_{i,\uparrow}^\dagger p_{i,\uparrow} - p_{i,\downarrow}^\dagger p_{i,\downarrow} \rangle,$$

and

$$j_{i,\perp}^{(1/2)} = \frac{1}{\sqrt{3}} \left( j_{i,x}^{(1/2)} + j_{i,y}^{(1/2)} + j_{i,z}^{(1/2)} \right),$$

as the component perpendicular to the honeycomb lattice. Note that the magnetic moment (in units of the Bohr magneton $\mu_{\text{B}}$) is given by $\mathbf{M} = -2\mathbf{j}^{(1/2)}$ at each site.

The initial state is the ground state obtained by the exact diagonalization method. Photoexcitation is introduced through the Peierls phase,

$$c^\dagger_{i,a,\sigma} c_{j,b,\sigma} \to \exp[-i(\boldsymbol{r}_j - \boldsymbol{r}_i) \cdot \boldsymbol{A}(t)] c^\dagger_{i,a,\sigma} c_{j,b,\sigma} \ .$$

Here, the intersite distance is set to be unity. The optical conductivity spectra are calculated for the ground state as in [48]. The spectrum on the high-energy side of the Mott gap (~1 eV) is discrete because of the finite-size effect. The system treated by the exact diagonalization method consists of six sites (i.e., three unit cells); thus, the finite size effect is apparent. In the thermodynamic limit, it should be continuous and corresponds to the excited states containing doubly occupied site(s) in terms of holes. The small oscillator strengths of the low-energy component (< 0.5 eV) would be enhanced by phonons [32] which are not considered in this calculation. Since this compound has two sites in a unit cell, both optically allowed and optically forbidden transitions could be constructed, without the help of phonons.

For photo-induced dynamics, we use a pulse of duration $t_{\text{off}} = 2n\pi/\omega$ with $n$ being an integer and frequency ω,

$$A_a(t) = \theta(0 < t < t_{\text{off}}) \frac{-F_L - F_R}{\omega} \left[\sin\left(\omega t - \frac{\pi}{2}\right) + 1\right],$$

$$A_b(t) = \theta(0 < t < t_{\text{off}}) \frac{F_L - F_R}{\omega} \left[\cos\left(\omega t - \frac{\pi}{2}\right)\right].$$

Here, $A_a$ and $A_b$ stand for the components of the vector potential $\boldsymbol{A}$(t), which is shown above in the formula for the Peierls phase, along the $a$- and $b$-axes shown in Fig. 1(a). The photon helicity is described by $F_L$ and $F_R$. They are the electric field amplitudes of the light fields of left-hand and right-hand circular polarizations. We use $t_{\text{off}}$ = 96 fs for ω = 0.3 eV ($n$ = 7) and ω = 0.6 eV ($n$ = 14). Because the intersite distance is 3.45 Å [25], $F_R$ = 0.2, $F_L$ = 0.0 (σ+) or $F_R$ = 0.0, $F_L$ = 0.2 (σ-) corresponds to 5.8 MV/cm. The time-dependent Schrödinger equation is numerically solved as before [49]. The decay component is insignificant in Fig. 4(c) because no relaxation process is considered and the system is too small to act as a heat bath for this field amplitude.

Supplementary note 11

Details of theoretical analysis: High-frequency expansion in Floquet theory

In general, we need the non-resonant condition to avoid thermalization. (If the non-resonant condition were not satisfied, the system would be thermalized after the application of a continuous wave and its temperature would finally go up to infinity). The frequencies used in the present experiment and theory are below the Mott gap; thus, harmful doubly-occupied sites are hardly produced. Furthermore, the pulse duration is sufficiently short to avoid thermalization. Therefore, quantum Floquet theory is applicable to the present situation.

Thus, continuous waves are considered here:

$$A_a(t) = \frac{F_{L(R)}}{\omega} \sin \omega t, \quad A_b(t) = \mp \frac{F_{L(R)}}{\omega} \cos \omega t$$

With the use of $J_m(ij) \equiv J_m\left(\frac{F_{L(R)}}{\omega}\right) e^{\mp im\phi_{ij}}$, where $J_m(x)$ on the right-hand side is the $m$th-order Bessel function and $\phi_{ij}$ is the angle between $\boldsymbol{r}_i - \boldsymbol{r}_j$ and a reference axis, $H_m$ in quantum Floquet theory is given by

$$H_m = -\sum_{ij} \vec{c}_i^\dagger \{\mathbf{T}_{ij} J_m(ij) \otimes \mathbb{I}_{2\times 2}\} \vec{c}_j .$$

In the second-lowest order of the high-frequency expansion, we have [50]

$$H_F^{(2)} = \sum_{m>0} \frac{[H_m, H_{-m}]}{m\omega} .$$

Since the hopping in $H_{\text{hop}}$ is limited to the nearest neighbors [$\boldsymbol{X_1}, \boldsymbol{Y_1}, \boldsymbol{Z_1}$ in Fig. 4(a)], $H_F^{(2)}$ represents hopping to the next-nearest neighbors within each sublattice [$\boldsymbol{X_2}, \boldsymbol{Y_2}, \boldsymbol{Z_2}$ in Fig. 4(a)]. We estimate it by approximating the $\boldsymbol{k}$-dependent terms by $\boldsymbol{k} = 0$ and keeping the $m = 1$ terms only,

$$H_F^{(2)} \cong \frac{1}{\omega} \sum_{iab\sigma} J_1^2\left(\frac{F_{L(R)}}{\omega}\right) (\pm 2i) \sin \frac{2\pi}{3} ([\mathbf{T}_1^Y, \mathbf{T}_1^Z] + [\mathbf{T}_1^Z, \mathbf{T}_1^X] + [\mathbf{T}_1^X, \mathbf{T}_1^Y])_{ab} c_{i,a,\sigma}^\dagger c_{i,b,\sigma} .$$

Assuming the threefold symmetry, $t_1 = t'_{1a} = t'_{1b}$, $t_2 = t'_2$, $t_3 = t'_3$, $t_4 = t'_{4a} = t'_{4b}$, we have

$$H_F^{(2)} \cong \frac{1}{\omega} J_1^2\left(\frac{F_{L(R)}}{\omega}\right) (\pm\sqrt{3})(t_2 - t_4)[t_2 - t_4 + 2(t_3 - t_1)]$$

$$\times \sum_{i\sigma} \begin{pmatrix} c_{i,yz,\sigma}^\dagger & c_{i,xz,\sigma}^\dagger & c_{i,xy,\sigma}^\dagger \end{pmatrix} \begin{pmatrix} 0 & -i & i \\ i & 0 & -i \\ -i & i & 0 \end{pmatrix} \begin{pmatrix} c_{i,yz,\sigma} \\ c_{i,xz,\sigma} \\ c_{i,xy,\sigma} \end{pmatrix},$$

which implies the application of an effective magnetic field to the effective angular momenta in the direction perpendicular to the honeycomb lattice. Because of the inequality $(t_2 - t_4)[t_2 - t_4 + 2(t_3 - t_1)] < 0$ owing to the opposite signs of the dominant transfer integrals $t_2$ and $t_3$, the effective magnetic field originating from $H_F^{(2)}$ points to the direction of $(1,1,1)$ for left-hand circular polarization and to the direction of $(-1,-1,-1)$ for right-hand circular polarization. Note that this effective magnetic field emerges without relying on the spin-orbit coupling.